\documentclass[10pt,conference,a4paper]{IEEEtran}
\usepackage{cite}
\usepackage{amsmath,amssymb,amsfonts}
\usepackage{algorithmic}
\usepackage{graphicx}
\usepackage{textcomp}
\usepackage{xcolor}
\usepackage{booktabs}
\usepackage{multirow}
\usepackage{hyperref}
\usepackage{enumerate}
\usepackage{float}
\usepackage[normalem]{ulem} 

\newif\ifredline
\redlinefalse
\definecolor{rladd}{RGB}{0,140,0}
\definecolor{rldel}{RGB}{210,0,0}
\makeatletter
\DeclareRobustCommand{\add}[1]{\ifredline\leavevmode{\color{rladd}#1}\else#1\fi}
\DeclareRobustCommand{\del}[1]{\ifredline\leavevmode{\color{rldel}\sout{#1}}\else\@bsphack\@esphack\fi}
\DeclareRobustCommand{\rep}[2]{\ifredline\leavevmode{\color{rldel}\sout{#1}}\ {\color{rladd}#2}\else#2\fi}
\makeatother

\begin{document}

\title{Documentation Drift in AI-Generated Pull Requests}

\author{
\IEEEauthorblockN{Will De Michele}
\IEEEauthorblockA{%
\textit{Monash University} \\
Melbourne, Australia 
}
\and
\IEEEauthorblockN{Kla Tantithamthavorn}
\IEEEauthorblockA{%
\textit{Monash University} \\
Melbourne, Australia 
}
}

\maketitle

\begin{abstract}
AI coding agents now contribute to software repositories at scale, yet their documentation practices remain uncharacterised. We analyse 29,669 AI-generated pull requests from \rep{1,513}{2,585} repositories in the AIDev dataset, measuring documentation co-change: whether code-file modifications are accompanied by a documentation\add{-file} update within the same PR. We find that 63.2\% of AI PRs change code without co-updating \add{any }documentation, varying sharply by task type (feature 55.2\% vs. refactor 77.5\%). \rep{This drift carries measurable downstream costs. Undocumented code files are re-fixed}{Drift is associated with later fixes: undocumented code files are followed by a same-file fix} 8.6--11.7\% more often than documented ones across 30--90 day windows. (A raw PR-level comparison shows the opposite \del{only }because documented PRs are 3.9$\times$ larger and so attract more fixes; controlling for size exposure reverses it.) Documented changes also \rep{clear}{merge} faster despite their larger size (median 40s vs.\ 184s\del{ to merge})\rep{, consistent with documentation making a change's intent clear to the merge process}{, although merging here is largely automated}. \rep{The cost compounds: a}{Drift also compounds: a} file changed without documentation is \add{later }re-documented at roughly half the rate of one changed with documentation (24\% vs.\ 48\% within 90 days), so documentation debt \rep{accumulates rather than being repaid}{tends to accumulate rather than being revisited}. 
\end{abstract}


\section{Introduction}

Documentation is widely recognised as a basis of software maintenance~\cite{khamis2010automatic, silva2023towards}. When code evolves without corresponding updates to its documentation, the resulting inconsistency, termed \emph{documentation drift}, may increase defect rates, raise maintenance costs, and hinder developer comprehension~\cite{khamis2010automatic, silva2023towards}. Co-changing code and documentation within a single contribution mitigates this risk, keeping knowledge artefacts aligned with the codebase.

AI coding agents (GitHub Copilot, Cursor, Devin, Claude Code, OpenAI Codex) now autonomously generate code and create pull requests (PR) at scale. While prior work has examined agentic code quality~\cite{yetistiren2022assessing}, security~\cite{fu2025security}, and testing~\cite{el2024using}, little is known about how these agents maintain documentation alongside the code they produce.\del{ As agentic AI evolves rapidly, understanding the current state of documentation co-change is essential for identifying where tooling and workflows can improve.}

We conduct a large-scale analysis of documentation drift in AI-authored pull requests: for each PR we measure whether code changes are accompanied by documentation updates  \add{(taking the absence of any documentation co-change as an observable proxy for drift)}, and how \rep{the resulting drift relates}{this relates} to resolution time and to \rep{defects that later surface}{later fix activity} in the same files.
Through an empirical study of 29,669 AI-generated PRs \rep{and 7,432 bug-fix PRs across 1,513}{across 2,585} repositories from the AIDev dataset~\cite{li2025aiteammates}, we answer three research questions:

\begin{enumerate}[{\bf RQ1)}]
\item {\bf What is the prevalence of documentation drift in AI-generated pull requests?} \\
63.2\% of AI-generated PRs change code without co-updating \del{the related }documentation, and this rate varies sharply by task type (feature 55.2\% vs.\ refactor 77.5\%)\add{; every agent drifts in 48--73\% of its PRs}\del{, underscoring the scale of documentation drift among current AI coding agents}.

\item {\bf \rep{Does documentation drift carry a downstream cost---in merge friction or in subsequent defects?}{Is drift associated with merge time or with subsequent same-file fixes?}}\\
Yes. Undocumented files are \rep{re-fixed}{followed by a same-file fix} 8.6--11.7\% \add{(relative) }more often than documented ones across 30--90 day windows, and documented changes \rep{clear}{merge} faster despite being larger. \del{(A raw PR-level count shows the reverse only because documented PRs are 3.9$\times$ larger and so attract more fixes.)}

\item {\bf \rep{Does documentation debt accumulate: when code is changed without documentation, is that gap subsequently repaid?}{Does drift persist: is code changed without documentation later re-documented?}} \\
Drift \rep{is self-perpetuating}{persists}: a file changed without documentation is re-documented at half the rate of one changed with documentation (24\% vs.\ 48\% within 90 days)\del{, so documentation debt accumulates rather than being repaid}.

\end{enumerate}


\section{Related Work}

\textbf{Code--documentation co-evolution and drift.}
Code--documentation co-evolution has been studied extensively in
human-authored projects. Fluri et al. ~\cite{fluri2007cochange,
fluri2009coevolution}
showed that most comment edits occur in the same revision as the associated
code, yet much new code remains undocumented. Later work frames code-documentation divergence as
\emph{drift}~\cite{romeo2024did} and links code--comment inconsistency to
elevated defect risk~\cite{radmanesh2024inconsistency}; surveys catalogue
comment-quality dimensions such as completeness and
freshness~\cite{rani2023decade}. We differ in two respects: we operate at the
level of whole \emph{documentation files} co-changed within a PR rather than
inline comments, and focus on AI-agent contributions.

\textbf{Empirical studies of AI coding agents.}
The AIDev dataset catalogues hundreds of thousands of agent-authored PRs across
five agents and tens of thousands of repositories~\cite{li2025aiteammates}, and
a growing literature uses it to characterise agent behaviour: acceptance rates
depend strongly on task type~\cite{pinna2026comparing}; general agentic-coding
usage~\cite{watanabe2026useagenticcodingempirical}, human--AI differences in PR
structure~\cite{gonzalez2025different}, and security-related
PRs~\cite{siddiq2026securityageaiteammates} have all been examined. Closest to our study, Haque
et al. measure how often agents include \emph{tests} alongside
code~\cite{haque2026tests}. Beyond AIDev, work on AI-generated code has assessed
functional quality~\cite{yetistiren2022assessing},
security~\cite{fu2025security}, and
testing~\cite{el2024using}. To our knowledge no prior study quantifies
documentation \emph{co-change} within AI-authored PRs or relates it to
subsequent \rep{defects}{fixes}; we provide the first such analysis, a documentation
counterpart to the test co-change of~\cite{haque2026tests}.


\section{Research Methodology}

\del{Our study proceeds in three stages. We first assemble a set of AI-generated PRs
with commit-level file details (\mbox{Section~\ref{sec:data}}); we then identify, for
each PR, whether it co-changes documentation alongside code
(\mbox{Section~\ref{sec:identification}}); and finally we define the maintainability
metrics (resolution time, subsequent bugs, and documentation-debt
repayment) through which we relate documentation drift to downstream outcomes
(\mbox{Section~\ref{sec:metrics}}).}

\subsection{Data Collection}
\label{sec:data}

We draw on the AIDev dataset (MSR 2026 Mining Challenge)~\cite{li2025aiteammates},
a public corpus of pull requests authored by five AI coding agents
(Claude\_Code, Cursor, Devin, GitHub Copilot, and OpenAI\_Codex) across tens of
thousands of open-source repositories between 24~December~2024 and
4~August~2025. Our analysis requires \emph{file-level} change information, so we
build on the \texttt{pr\_commit\_details} table, which records each modified file
per commit with its added and deleted line counts; we aggregate these to one row
per (PR, file), summing line changes across a PR's commits.

We retain a PR only if it \rep{(i) was merged, (ii) has commit-level file details
available, and (iii)}{(i) has commit-level file details available and (ii)} modifies at least one code file (defined in
Section~\ref{sec:identification}). PRs that touch no code file (e.g.\
documentation-only or asset-only PRs) are excluded, as documentation co-change
is undefined for them. Applying these criteria yields \textbf{29,669 AI-generated
PRs  \add{(merged or not)} across \rep{1,513}{2,585} repositories}\rep{, comprising 177,743 individual code-file
changes. Within this set, 7,432 PRs are bug-fix PRs (task type
fix; \mbox{Section~\ref{sec:metrics}}), used for
file-level bug attribution.}{, on which we measure prevalence. All other analyses use the \textbf{19,881 merged PRs} (1,513 repositories; \textbf{177,743 code-file changes}), whose 4,526 bug-fix PRs (task type \texttt{fix}) supply the fix events (Section~\ref{sec:metrics}).} 

\add{Our replication package (classification and analysis scripts) is available at \url{https://doi.org/10.5281/zenodo.21873884}.}

\subsection{Documentation Co-Change Identification}
\label{sec:identification}

\rep{To answer RQ1 we identify, for each PR, whether documentation was updated
alongside code within the same contribution. This is a two-step, deterministic mapping from file paths to categories.}{We label each PR in two deterministic steps from its file paths.}

\textbf{Step 1: File-type identification.} We assign every changed file to one of
three categories from its path, extension, and filename:
\begin{itemize}
\item \textbf{Code}: source code (\texttt{.py}, \texttt{.js}, \texttt{.java}, \ldots), configuration (\texttt{.json}, \texttt{.yaml},
\texttt{.toml}), and build files (\texttt{.tf}, \texttt{.sh},
Dockerfile, Makefile).
\item \textbf{Documentation}: Markdown and text formats (\texttt{.md},
\texttt{.mdx}, \texttt{.rst}, \texttt{.txt}), any file under a \texttt{docs/}
directory, and conventional documentation filenames (\texttt{README},
\texttt{LICENSE}, \texttt{CHANGELOG}).
\item \textbf{Other}: everything else, including images and media assets and
machine-generated or binary artefacts. These files do not enter
the co-change decision.
\end{itemize}

\textbf{Step 2: PR-level labelling.} Using the code/documentation split, we label
each PR as either \emph{co-changed} or \emph{drifted}:
\begin{itemize}
\item \textbf{Co-changed}: the PR modifies
$\geq$1 code file \emph{and} $\geq$1 documentation file.
\item \textbf{Drifted}: the PR modifies
$\geq$1 code file and \emph{0} documentation files.
\end{itemize}
We use \emph{co-change}/\emph{drift} for the two phenomena and
\emph{co-changed}/\emph{drifted} for the corresponding PR labels throughout.
\add{Drift is thus a proxy,it flags a missing documentation update, not a verified inconsistency between code and its related documentation (Section~\ref{sec:threats}).}

\textbf{Task type.} PRs are labelled in the dataset with the taxonomy (\texttt{feat},
\texttt{fix}, \texttt{docs}, \texttt{refactor}, \texttt{chore}, \ldots).

\subsection{Metrics}
\label{sec:metrics}

\del{We study three aspects of documentation drift: the friction a PR meets before it
merges, the defects that later surface in the files it changed, and whether the
gap it leaves behind is ever repaid.}

\textbf{Resolution time and friction (PR level).} We measure elapsed time from PR
creation to merge. The distribution is right-skewed (most PRs merge almost
instantly, a minority stay open for hours or days), so we report median and mean,
plus the share merged within five minutes (isolating effectively automated
merges). We add two friction
counts from the PR timeline: comments and \emph{review touches} (reviews + label
changes + review requests).

\textbf{\rep{Bug debt}{Subsequent same-file fixes} (file-level attribution).} We count subsequent bug-fix PRs
\rep{attributable to}{that touch} the files a PR changed, within 30-, 60-, and 90-day windows. From
\add{merged }bug-fix PRs (task type \texttt{fix}) we derive \emph{fix events}, one per
(repository, file, merge time). For a code file changed in a PR, a subsequent \rep{bug}{fix}
is a fix event on the same repository and file path whose merge time falls
strictly after the PR's, within the window (a PR never attributes a \rep{bug}{fix} to
itself). We record per-file counts (summed to PR level) and the binary ``at least
one subsequent fix'' outcome. \add{Each fix event is credited to every earlier PR that changed the file within the window, so one fix may count for several PRs and a file recurs once per PR that changes it; we report file-level rates descriptively, without independence-based tests.} This attribution is temporal, not causal, a
limitation we revisit in \mbox{Section~\ref{sec:threats}}. \add{We do not restrict to PRs with a complete window: AIDev ends on 4~August~2025, so PRs merged near the cutoff have truncated (right-censored) follow-up, lowering observed rates, especially at 60 and 90 days.}

\textbf{Exposure control.} Larger PRs touch more files, offering more surface for
later fixes to attach to and confounding PR-level \rep{bug}{fix} comparisons with change
size. We address this two ways. (1)~We compare co-changed and drifted PRs only
within \rep{identical file-count bins}{bins of equal code-file count}. (2)~Since binning is coarse, we shift the unit
of analysis to the individual changed code file (177,743 file-changes), holding
exposure fixed at a single file\del{ so that no change gains bug surface merely by
touching more}. At this granularity we report, per window, the probability that a
changed code file is later \rep{re-fixed}{touched by a fix PR}, split by whether its PR co-changed
documentation, and \rep{check that the pattern holds}{examine the pattern} within per-file line-change bins,
which hold individual file size approximately fixed.

\textbf{\rep{Documentation-debt repayment}{Re-documentation}.} To test whether drift \rep{accumulates rather
than being corrected}{persists}, we track each changed code file forward and ask whether any
later PR touching it co-changes documentation within 90~days (``\rep{repayment}{re-documentation}''),
comparing files changed \emph{without} documentation against those changed
\emph{with} it. \add{This captures renewed documentation activity on the file, not verified repayment of the original gap, since the later documentation change may concern other code.} We quantify this stickiness across repositories
(repository-clustered logistic regression controlling for per-file lines changed
and agent) and within repositories (repository fixed-effects linear-probability
model over the 487 repositories exhibiting both behaviours), the latter isolating
a file-level dynamic from repository documentation culture.

\section{Results}

\subsection{RQ1: Prevalence of Documentation Drift}

\textbf{Prevalence:} Of the 29,669 AI-generated PRs analysed, 18,751 (63.2\%) do not include a documentation co-change alongside their code modifications, while 10,918 (36.8\%) do co-change documentation. This suggests a substantial gap in documentation co-change frequency among current AI coding agents. \add{Lacking a human baseline, we do not claim this rate is specific to agents.}

\begin{figure}[h]
  \centering
  \includegraphics[width=0.75\linewidth]{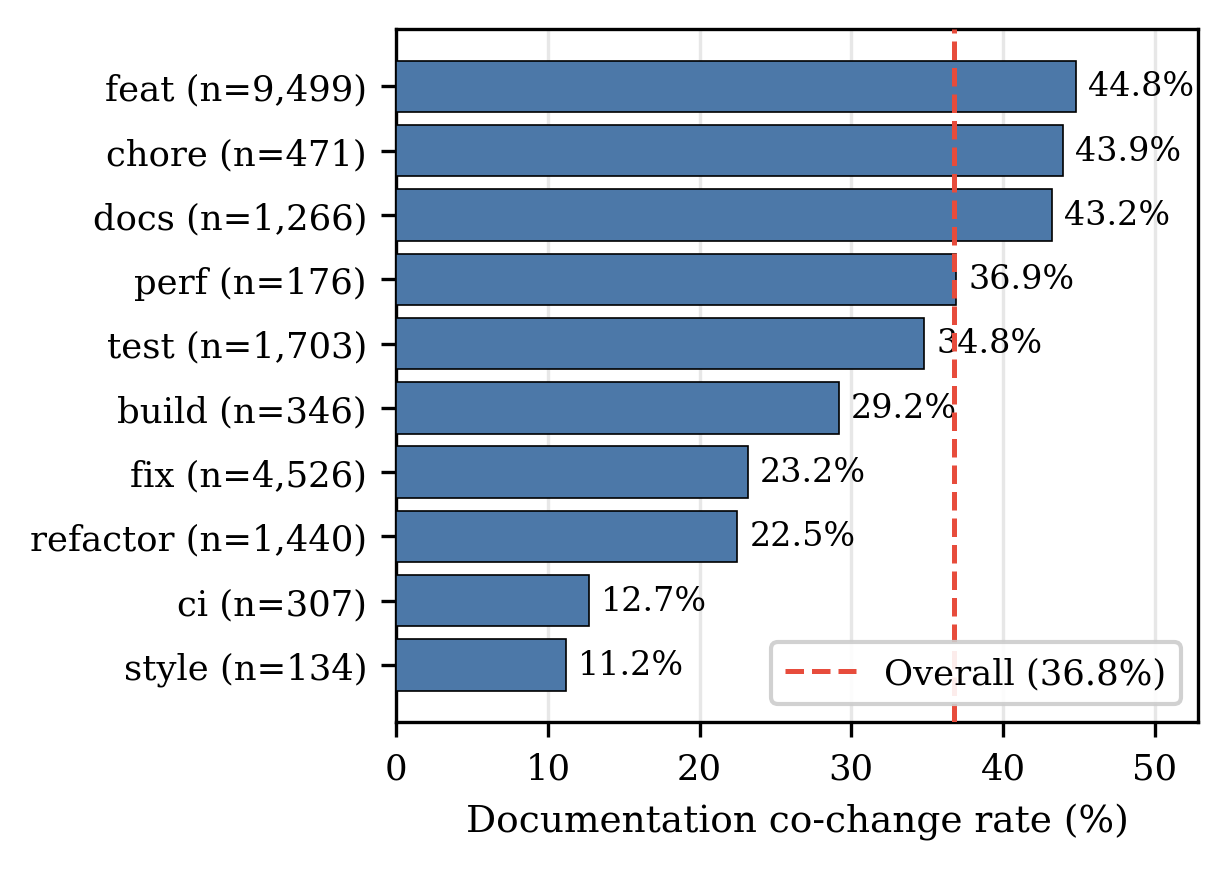}
  \caption{Documentation co-change rate by task type (\add{merged PRs; }task types with $n\geq100$ PRs), with overall co-change rate \add{over all 29,669 PRs }(36.8\%).}
  \label{fig:task-consistency}
\end{figure}

As shown in Figure~\ref{fig:task-consistency}, co-change rates vary by task type: feature PRs exhibit the highest co-change rate (44.8\%), followed by chore (43.9\%) and docs (43.2\%), while fix (23.2\%) and refactor (22.5\%) PRs are \add{among the }least likely to include documentation co-changes. This pattern suggests that agents are more likely to co-update documentation for additive tasks than for corrective or restructuring work\add{, although internal fixes and refactorings may also less often need a documentation change}.

\begin{figure}[h]
  \centering
  \includegraphics[width=0.75\linewidth]{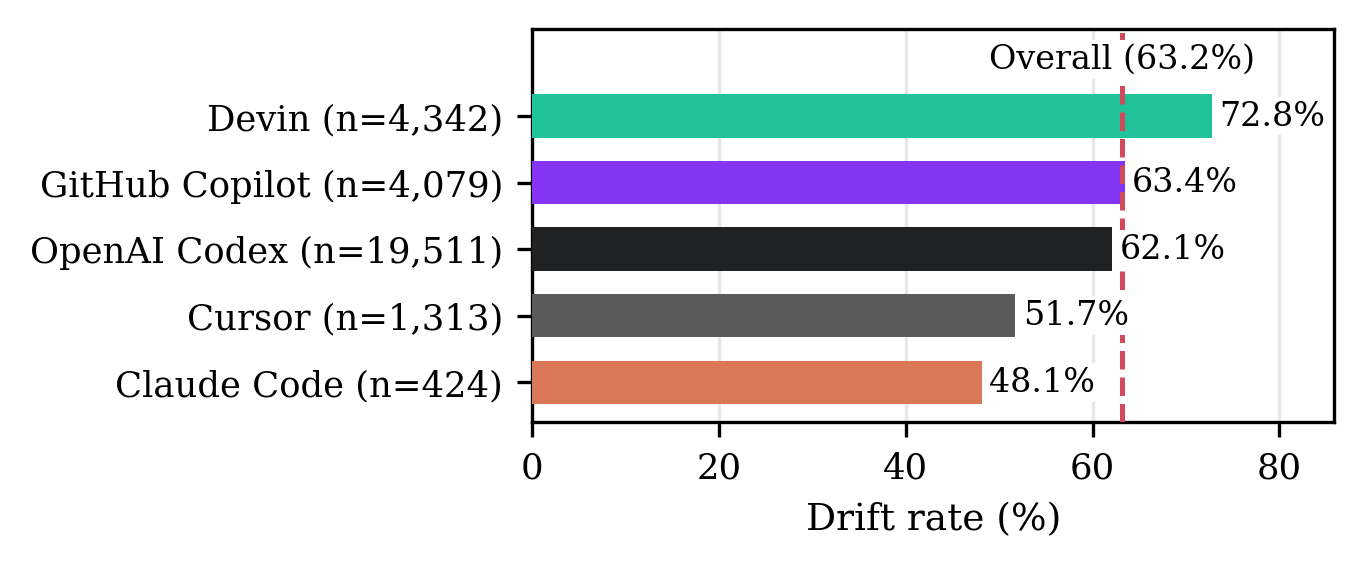}
  \caption{\add{Drift rate by agent (all 29,669 PRs), overall drift rate (63.2\%).}}
  \label{fig:agent}
\end{figure}

\add{\textbf{By agent:} Drift is widespread across agents rather than concentrated in one (Figure~\ref{fig:agent}): it ranges from 48.1\% for Claude Code to 72.8\% for Devin, and is 65.3\% without Codex, which contributes 65.8\% of the PRs. Nor is the task-type contrast an artefact of agent mix: within every agent, feature PRs co-change documentation 1.6--2.1$\times$ as often as fix PRs.}

\textbf{Answer:} Documentation drift is prevalent: 63.2\% of AI PRs \add{(48--73\% per agent)} change code without co-updating documentation, and co-change tracks task type \add{within every agent}: agents document additive work more than corrective or restructuring work.\del{ Whether this drift carries any downstream cost is the question we turn to in RQ2.}

\subsection{\texorpdfstring{RQ2: \rep{Does Documentation Drift Carry a Downstream Cost?}{Is Drift Associated with Merge Time or Later Fixes?}}{RQ2: Is Drift Associated with Merge Time or Later Fixes?}}

\rep{Drift is prevalent, but does it cost anything? We examine two channels: friction at merge time (resolution time) and subsequent defects in the same files. We find that drift carries a cost on both: undocumented changes are re-fixed more often, and documented changes clear faster.}{We examine merge time and subsequent same-file fixes; drift is associated with both, though neither association establishes causation.}

\textbf{Documented changes \rep{clear}{merge} faster despite being larger.} Resolution time in this dataset is shaped by automated merging (most AI \rep{PR reviews}{PRs} merge within minutes) so it is a noisy proxy for the human effort a contribution demands, and we read its direction as suggestive rather than conclusive. \del{Even so, the direction is informative.} As Table~\ref{tab:resolution} shows, co-changed PRs (which we later establish are $3.9\times$ larger)\del{,} nonetheless clear \emph{faster}: a lower median time to merge (40s vs.\ 184s) and a higher share merged within five minutes (67.3\% vs.\ 55.9\%).\del{ Change size normally slows merges, so a larger class of changes that clears faster is striking.}

\begin{figure}[h]
  \centering
  \includegraphics[width=0.55\linewidth]{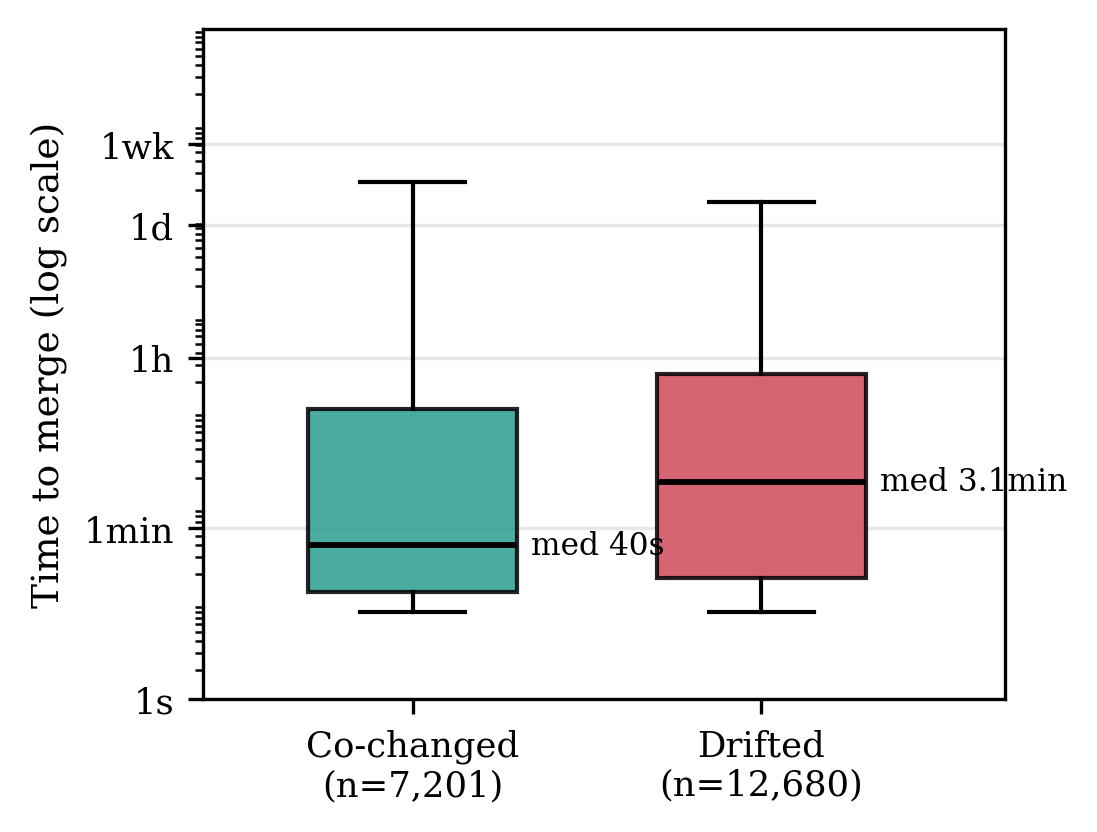}
  \caption{Time to merge for co-changed vs.\ drifted PRs. Documented (co-changed) changes \rep{clear}{merge} faster despite being $3.9\times$ larger\del{, consistent with documentation aiding comprehension even in an automation-heavy workflow}.}
  \label{fig:resolution-time}
\end{figure}

\begin{table}[h]
\centering
\caption{Resolution Time: Co-changed vs Drifted (merged AI PRs)}
\label{tab:resolution}
\begin{tabular}{lrr}
\toprule
\textbf{Metric} & \textbf{Co-changed} & \textbf{Drifted} \\
\midrule
Median resolution & 40s & 184s \\
Mean resolution & 17.8h & 10.0h \\
Merged within 5 min & 67.3\% & 55.9\% \\
Comments (mean) & 0.98 & 1.03 \\
Review touches (mean) & 1.94 & 1.76 \\
\bottomrule
\end{tabular}
\end{table}

The faster clearance is not explained by lighter review: review touches are, if anything, slightly \emph{higher} for co-changed PRs (1.94 vs.\ 1.76). The mean time to merge is longer for co-changed PRs (17.8h vs.\ 10.0h), reflecting a heavier tail rather than a slower typical case. \rep{The faster median most likely reflects documentation making}{One plausible explanation for the faster median is that documentation makes} a change's intent clearer, to reviewers and to the automated pipelines that dominate this workflow, so documented changes \rep{clear}{merge} more readily than their size alone would predict, while drifted changes \add{may }take longer to understand. \add{We cannot test this here: merge speed also depends on task type, agent, repository workflow and automation policy, which we do not control.}

\textbf{The PR-level \rep{bug}{fix} view is misleading.} Table~\ref{tab:bugs} shows that, at the PR level, documented (co-changed) PRs are surprisingly followed by 79\% \emph{more} file-attributed \rep{bugs}{fixes} within a 30-day window (mean 3.78 vs.\ 2.11, $p<0.001$; 60\% vs.\ 40\% followed by at least one such \rep{bug}{fix}). \del{On its own, this would appear to count against documentation.} But co-changed PRs are also 3.9$\times$ larger (24.83 vs.\ 6.43 files, $p<0.0001$), with 66\% of the size gap in code files (16.34 vs.\ 4.17), and a larger PR simply offers more files for later fixes to attach to. \rep{The bug gap is therefore an exposure artefact of change size, not a property of documentation}{The gap therefore largely reflects exposure to change size rather than documentation}; and, as we now show, it inverts once exposure is removed.

\begin{table}[h]
\centering
\caption{\rep{Bug Debt}{Subsequent Same-File Fixes per PR} (30-day file-level attribution). $\Delta = (\text{Co-changed} - \text{Drifted})/\text{Drifted}$.}
\label{tab:bugs}
\begin{tabular}{lrrr}
\toprule
\textbf{Metric} & \textbf{Co-changed} & \textbf{Drifted} & \textbf{$\Delta$} \\
\midrule
\rep{Mean bugs}{Mean fixes} & 3.78 & 2.11 & +79\% \\
\rep{Median bugs}{Median fixes} & 1.0 & 0.0 & - \\
Files changed & 24.83 & 6.43 & +286\% \\
Code files & 16.34 & 4.17 & +292\% \\
\rep{Bugs/file}{Fixes/file} & 0.85 & 1.06 & -20\% \\
\bottomrule
\end{tabular}
\end{table}

\begin{figure}[h]
\centering
\includegraphics[width=\linewidth]{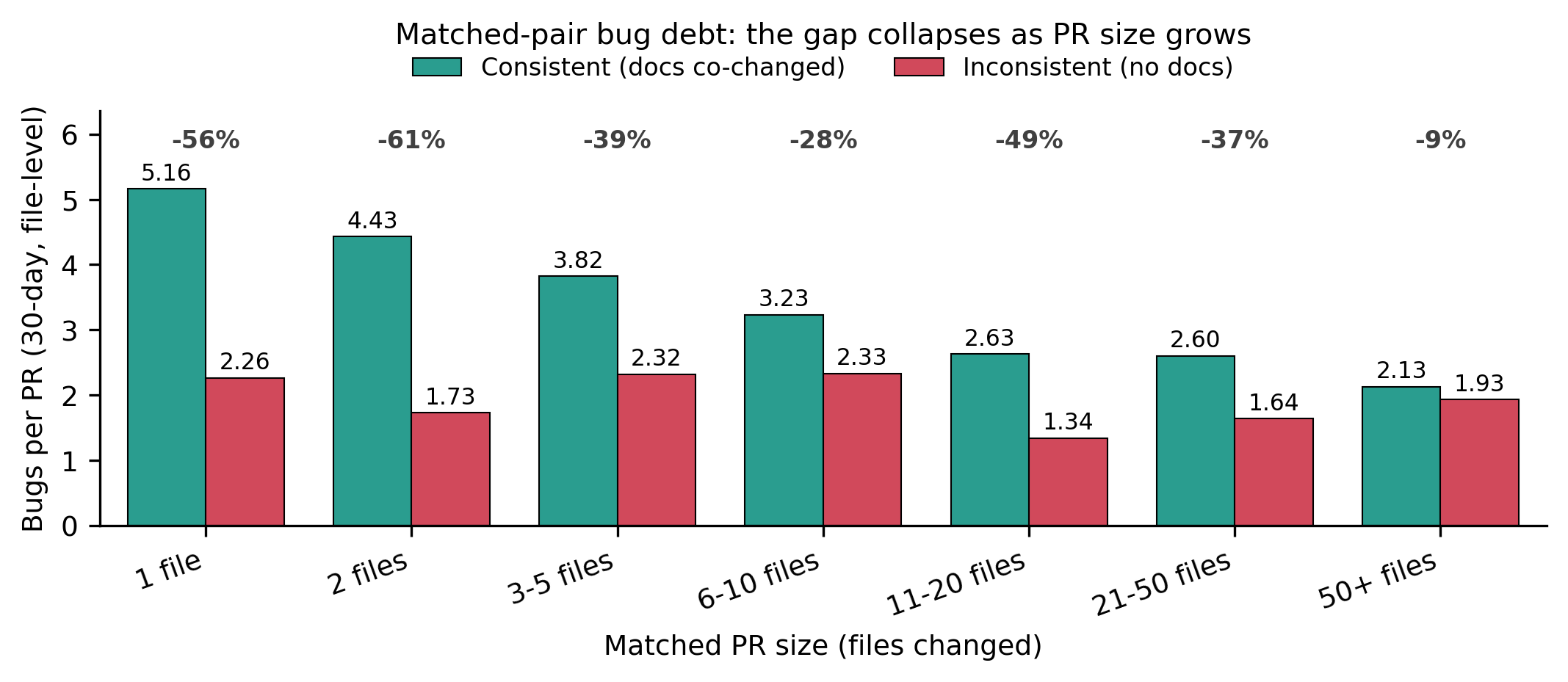}
\caption{\rep{Matched-Pair Bug Debt (30-day file-level). The PR-level co-change–drift gap collapses monotonically as PR size grows, the pattern expected from of an exposure artefact that inflates the apparent bug count of the larger, documented PRs.}{Same-file fixes per PR within 30 days, by code-file count. Co-changed PRs have more fixes in every bin, but the gap shrinks overall as PRs grow.}}
\label{fig:matched}
\end{figure}

\textbf{At the file level, drift is \rep{the defect-prone state}{associated with more later fixes}.} \rep{File-count matching does not remove the confound. Co-changed PRs show more bugs in every size bin (\mbox{Figure~\ref{fig:matched}}), the pattern expected from an exposure effect,}{Binning by code-file count does not remove the confound: co-changed PRs still show more fixes in every bin (Figure~\ref{fig:matched}), as the PR-level count also matches fixes through the documentation files they also touch,} so we move the unit of analysis to the individual changed code file, holding exposure fixed at one file (177,743 file-changes across 1,513 repositories). At this granularity the signal \emph{reverses}: an undocumented (drifted) code file is \rep{re-fixed}{followed by a same-file fix} \emph{more} often than a documented one, by 8.6\% at 30 days, 11.3\% at 60, and 11.7\% at 90 (\add{relative; 1.2--1.7 percentage points; }Table~\ref{tab:filebug}). \rep{The gap widens with the window, as expected if drift-related defects surface gradually rather than at once, and it}{The gap} is not an artefact of a few large files: within every per-file line-change bin below 50 lines (covering the large majority of changes) drifted files are \rep{re-fixed}{followed by a fix} more often (e.g.\ 15.6\% vs.\ 11.9\% for 1--5 line changes). \del{Drift, not documentation, is the defect-prone state.}

\begin{table}[h]
\centering
\caption{File-level \rep{re-fix probability}{probability of a subsequent same-file fix} (exposure fixed at one changed code file).}
\label{tab:filebug}
\begin{tabular}{lrrr}
\toprule
\textbf{Window} & \textbf{Documented} & \textbf{Drifted} & \textbf{Drift excess} \\
\midrule
30 days & 14.1\% & 15.3\% & +8.6\% \\
60 days & 14.8\% & 16.5\% & +11.3\% \\
90 days & 14.9\% & 16.6\% & +11.7\% \\
\bottomrule
\end{tabular}
\end{table}

\textbf{Answer:} \rep{Drift carries a downstream cost: undocumented files are re-fixed}{Drift is associated with more subsequent fixes: undocumented files are followed by a same-file fix} 8.6--11.7\% more often and documented changes \rep{clear}{merge} faster despite being $3.9\times$ larger; the PR-level \rep{bug}{fix} count points the other way \del{only }because size confounds it.

\subsection{\texorpdfstring{RQ3: \rep{Does Documentation Debt Accumulate?}{Does Drift Persist?}}{RQ3: Does Drift Persist?}}

\begin{figure}[h]
  \centering
  \includegraphics[width=0.4\linewidth]{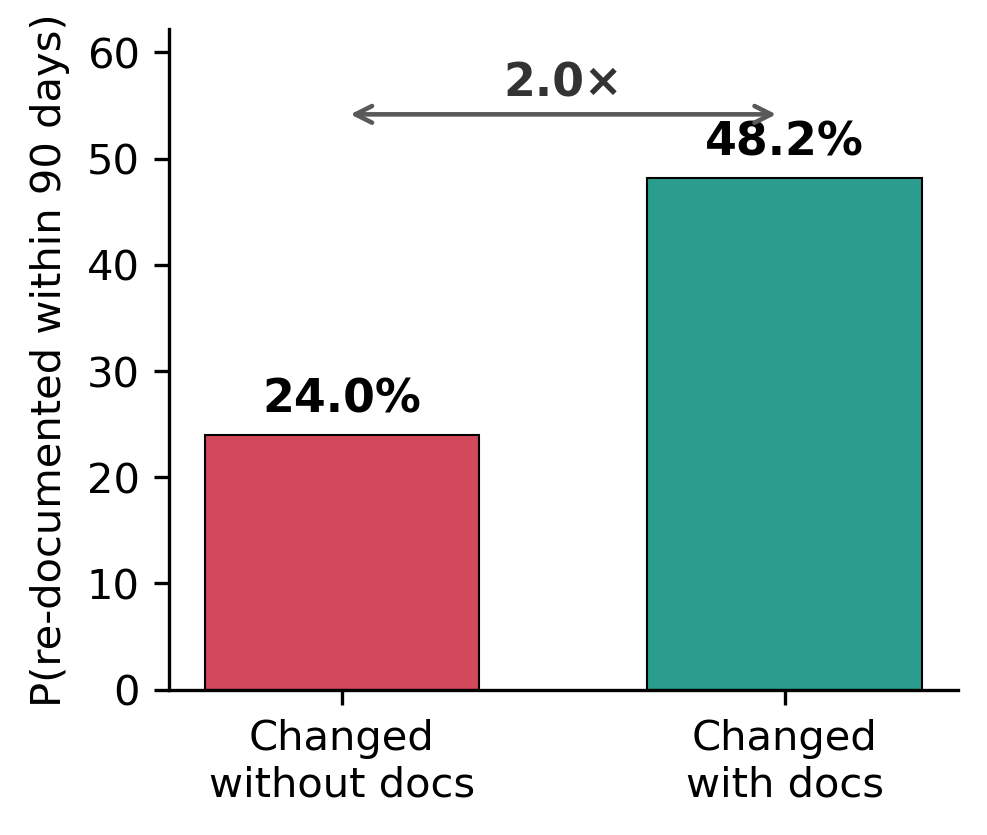}
  \caption{\rep{Documentation debt is perpetuating: code}{Code} files changed without documentation are re-documented \add{within 90 days }at half the rate of \add{those }changed with documentation.}
  \label{fig:stickiness}
\end{figure}

The RQ2 cost is per-change; RQ3 asks whether drift also compounds. \del{It does: documentation debt is self-perpetuating.} We follow each changed code file forward and ask whether it is subsequently \add{re-}documented (a later PR touching the same file that co-changes documentation) within 90 days. A file changed \emph{without} documentation is re-documented only 24.0\% of the time, versus 48.2\% for a file changed \emph{with} documentation (Figure~\ref{fig:stickiness}), so the initial choice to skip documentation tends to persist\del{ rather than being repaid}. Controlling for per-file lines changed and agent, with errors clustered by repository, a currently-documented file has 3.2$\times$ the odds of later documentation (95\% CI 2.20--4.69, $p<10^{-9}$). Because documenting repositories differ systematically from non-documenting ones, we also fit a repository fixed-effects model restricted to the 487 repositories that exhibit both behaviours: \emph{within the same repository}, documenting a file raises its probability of later documentation by 18.9 percentage points (95\% CI 3.0--34.8, $p=0.02$). The effect shrinks but survives, \rep{indicating a genuine file-level dynamic rather than mere repository culture}{suggesting a file-level pattern beyond repository culture}. \add{Both rates are lower bounds, as 90-day windows near the cutoff are censored.}

\textbf{Answer:} Drift \rep{accumulates}{persists}: undocumented code is re-documented at roughly half the rate of documented code, and the stickiness holds within repositories\del{: a second, self-reinforcing cost on top of the per-file defect cost of RQ2}.

\section{Discussion}
\label{sec:discussion}

\textbf{\rep{Undocumented changes are the riskier ones.}{Drift as a risk marker.}} Code changed without documentation is \rep{re-fixed}{followed by a same-file fix} more often than code changed with it\del{, and documented changes also clear faster than their size predicts}. Drift \rep{is}{may} thus \add{be} an indicator of defect-prone change: the file that skipped documentation \rep{is the one}{tends to be the one} more likely to come back. We cannot settle \emph{why} from these data (documentation may aid the agent's own reasoning \rep{of}{about} a change, \rep{or its absence may mark hurried, less-considered edits), but the two signals point the same way, so documentation co-occurs with higher-quality contributions}{its absence may mark hurried edits, or the association may reflect task type, agent or repository differences), nor rule out that the later fixes address older or unrelated defects}.

\textbf{\rep{The cost compounds.}{Drift persists.}} \rep{Drift is not only more defect-prone in the files it touches (RQ2); it is also self-perpetuating (RQ3)}{Beyond its association with later fixes (RQ2), drift is sticky (RQ3)}: once an agent changes a file without documentation, that file is far less likely to be documented later, even within the same repository. The concern is not that any single undocumented change is catastrophic (most are not) but that agents, operating at high volume and merge velocity, lay down documentation debt faster than it is repaid. Over time this widens the gap between what the code does and what is written down, precisely the divergence the human co-evolution literature warns against~\cite{fluri2007cochange, romeo2024did}\del{, but accruing at agent speed}. \del{Because both the per-file defect cost and the accumulation are invisible to per-PR quality checks, current evaluation of AI contributions may systematically understate the cost of drift.}

\textbf{Recommendations.} \emph{Organisations:} trigger human review on change \emph{size} (e.g., $>$5 files), not on the presence of documentation, and treat drift itself as a risk signal, since undocumented changes are \rep{re-fixed more often and the gap compounds}{more often followed by fixes and drift persists}. \emph{Agent developers:} encourage agents to perform documentation co-change, especially \rep{for the fix and refactoring work where drift is highest}{when a change alters documented behaviour}, and prioritise \emph{repaying} accumulated debt by prompting agents to document previously-undocumented files they revisit, not only files they create. \emph{Researchers:} attribute \rep{defects}{fixes} at file granularity with explicit exposure control, since PR-level \rep{bug}{fix} counts are confounded by size and can invert the sign of the documentation \rep{effect}{association}.

\subsection{Threats to Validity}
\label{sec:threats}

\textbf{Construct.} Our label captures whether \add{any }documentation \add{file }is updated alongside code, not \add{whether an update was needed or related to the changed code, nor }its \emph{quality} or \emph{semantic correctness}: \rep{a trivial README edit counts as co-change}{an unrelated CHANGELOG edit counts as co-change, while an internal refactoring that needs no documentation counts as drift}. We also see only in-repo documentation \emph{files}: inline comments, docstrings, and out-of-repo docs (wikis, hosted sites, API portals) are invisible to a commit-level method and count as drift, \rep{making our 63.2\% rate an upper bound}{inflating our 63.2\% rate (unrelated documentation edits deflate it)}. \del{Because this bias is broadly uniform across PRs, it shifts the baseline rather than our relative findings.}

\textbf{Internal.} \rep{The primary threat is confounding between co-change and PR
complexity, for which file count is only a proxy. Residual confounding remains
possible (we do not measure complexity or coupling), so our proxies may not
capture every dimension of change difficulty.}{Beyond change size, our RQ2 comparisons do not control for task type (strongly related to co-change, Figure~\ref{fig:task-consistency}), agent, repository workflow and automation policy, prior file churn, or change complexity; RQ3 controls for size, agent and repository but not task type.} Second, \rep{bug}{fix} attribution is
file-level and temporal, not \del{necessarily }causal: a same-file \texttt{fix} need
not stem from \rep{the original PR}{or touch the original change}, and \rep{bugs}{the defect} may be pre-existing or introduced by
intervening PRs. \add{Third, right-censored follow-up windows (Section~\ref{sec:metrics}) may bias comparisons if co-changed and drifted PRs differ in timing.}

\textbf{External.} Findings derive from AIDev (Dec 2024--Aug 2025), which is becoming outdated as agents evolve, and has substantial agent imbalance (Codex 65.8\%, Claude Code 1.4\%), thus weighting aggregate rates toward Codex-style workflows\add{; drift is widespread across agents (RQ1), but RQ2 and RQ3 pool agents, so their associations may differ by agent}. The human PR data lacks file-level detail, precluding a human baseline for the 63.2\% rate. As agentic AI evolves rapidly, these results are best read as an empirical baseline rather than a fixed characterisation.

\section{Future Work}

A file-level \emph{human baseline} would establish whether the 63.2\% drift rate is high or low relative to human practice. \add{Manual validation of a PR sample, }\rep{Moving}{moving} beyond presence/absence to \emph{documentation quality} (does the doc change reflect the code?)\add{,} and broadening the unit of documentation to inline comments, docstrings, and out-of-repo sources would sharpen the construct. Finally, \emph{line-level \rep{bug}{fix} attribution}, tracing each subsequent fix to the
exact lines it repairs rather than inferring
it from same-file co-occurrence, would replace our temporal attribution with a
causal link, establishing whether defects genuinely originate in the drifted change.

\section{Conclusion}

We present the first large-scale study of documentation co-change in AI-generated pull requests. Across 29,669 PRs from \rep{1,513}{2,585} repositories, 63.2\% modify code without co-updating documentation. \rep{This drift is not benign: undocumented files are re-fixed}{Drift is consistently associated with worse outcomes: undocumented files are followed by a same-file fix} 8.6--11.7\% more often than documented ones \add{across all three windows}, and documented changes \rep{clear}{merge} faster despite being 3.9$\times$ larger\del{; drift is the defect-prone state}. \rep{The cost compounds: undocumented}{Drift also persists: undocumented} code is re-documented at roughly half the rate of documented code, \rep{so documentation debt accumulates}{suggesting that documentation debt accumulates} rather than being repaid. \rep{The downstream cost of AI-driven drift is real, compounding, and easily hidden}{This points to a hidden cost of AI-assisted development, one that may compound over time and is easily missed} by per-PR metrics that conflate documentation with change size. \add{Coding agents, and the benchmarks that evaluate them, should treat documentation co-change as a first-class signal rather than an optional by-product.}

\bibliographystyle{IEEEtran}
\bibliography{references.bib}

\end{document}